\documentclass[twocolumn,showpacs,prc,floatfix]{revtex4}
\usepackage{graphicx}
\usepackage[dvips,usenames]{color}
\usepackage{amsmath}
\newcommand{\bs}{\begin{sloppypar}} \newcommand{\es}{\end{sloppypar}}
\def\beq{\begin{eqnarray}} \def\eeq{\end{eqnarray}}
\def\beqstar{\begin{eqnarray*}} \def\eeqstar{\end{eqnarray*}}
\newcommand{\bal}{\begin{align}}
\newcommand{\eal}{\end{align}}
\newcommand{\beqe}{\begin{equation}} \newcommand{\eeqe}{\end{equation}}
\newcommand{\p}[1]{(\ref{#1})}
\newcommand{\tbf}{\textbf}

\newcommand{\tcal}{\cal}

\begin {document}
\title{Correlation functions for a di-neutron  condensate
 in asymmetric nuclear matter}
\author{ A. A. Isayev}
 \affiliation{Kharkov Institute of
Physics and Technology, Academicheskaya Street 1,
 Kharkov, 61108, Ukraine
\\
Kharkov National University, Svobody Sq., 4, Kharkov, 61077,
Ukraine
 }
 \date{\today}
\begin{abstract}
Recent calculations with an effective isospin dependent contact
interaction show the possibility of the crossover from
superfluidity of neutron Cooper pairs in $^1S_0$ pairing channel
to Bose-Einstein condensation (BEC) of di-neutron bound states in
dilute nuclear matter. The density and spin correlation functions
are calculated for a di-neutron condensate in asymmetric nuclear
matter with the  aim to find the possible features of the BCS-BEC
crossover. It is shown that the zero-momentum transfer spin
correlation function satisfies the sum rule at zero temperature.
In symmetric nuclear matter, the density correlation function
changes sign at low momentum transfer across the BCS-BEC
transition  and this feature can be considered as a signature of
the crossover.   At finite isospin asymmetry, this criterion gives
too large value for the critical asymmetry $\alpha_c^d\sim0.9$, at
which the BEC state is quenched.  Therefore,  it can be trusted
for the description of the density-driven BCS-BEC crossover of
neutron pairs  only at small isospin asymmetry.
 This result generalizes the conclusion of the study in
Phys. Rev. Lett. {\bf 95}, 090402 (2005), where the change of sign
of the density correlation function at low momentum transfer in
two-component quantum fermionic atomic gas with the balanced
populations of fermions of different species was considered as an
unambiguous signature of the BCS-BEC transition.
\end{abstract}
\pacs{21.30.Fe, 21.60.-n, 21.65.Cd} \keywords{Di-neutron
condensate, BCS-BEC crossover, density and spin correlation
functions, asymmetric nuclear matter} \maketitle

\section{Introduction}
Few-body correlations play an important role in low-density
nuclear systems such as  expanding nuclear matter in heavy ion
collisions, the tails of nuclear density distributions in exotic
nuclei, or nuclear matter in the crust of a neutron star. Feasible
low-density  phenomena include  $\alpha$-particle
condensation~\cite{GSSN}, the formation of two-body
(deuteron)~\cite{BLS} and three-body (triton)~\cite{BSKR} bound
states, and their possible coexistence/competition~\cite{SC}. Of
special interest also  is the crossover from superfluidity of
nucleon Cooper pairs to  Bose-Einstein condensation (BEC) of
two-body bound states in dilute nuclear matter. The transition
from large overlapping Cooper pairs to tightly bound pairs of
fermions can be described on the basis of BCS theory, if the
effects of fluctuations are disregarded. First, this transition
was studied in excitonic semiconductors~\cite{E}, ordinary
superconductors~\cite{L} and  in an attractive Fermi
gas~\cite{NS}.  Although the BCS and BEC limits are physically
quite different, the crossover between them was found to be smooth
within the BCS theory. In nuclear matter, the BCS-BEC crossover is
usually mentioned in respect to the formation of Bose-Einstein
condensate of deuterons at low densities, which is caused by
strong neutron-proton ($np$) pair correlations in the
$^3S_1-^3D_1$ pairing channel~\cite{BLS,AFRS}. During this
transition the chemical potential changes  sign at certain
critical density (Mott transition), approaching  half of the
deuteron binding energy at ultra low densities. Despite the fact
that in the BCS regime even small isospin asymmetry effectively
destroys a $np$ condensate~\cite{AFRS,SL,AIPY}, at low densities
Bose-Einstein condensate of deuterons is weakly affected by an
additional gas of free neutrons even at very large
asymmetries~\cite{LNS,IYB}.

There is a rather general reason to expect that di-neutron
correlations  in low-density nuclear matter should be of
significance as well. As is known, the bare nucleon-nucleon
interaction in the $^1S_0$  channel leads to a virtual state
around  zero energy characterized by the large negative scattering
length $a_{nn}\approx-18.5\,\mbox{\rm fm}$. This implies a very
strong attraction between two neutrons in the spin singlet state
that may lead to the formation of compact bound pairs of neutrons
in low-density  nuclear systems. Recent theoretical studies
confirm this suggestion. In particular, the spatial structure of
the neutron Cooper pair in low-density nuclear and neutron matter
was analyzed with the use of the bare G3RS force and the effective
D1 Gogny interaction~\cite{M}. It was shown that the size of the
Cooper pair becomes smaller than the average inter-neutron
distance at a wide range of neutron densities
$\varrho_n/\varrho_0\sim 10^{-4}-10^{-1}$, independently of the
type of the interaction ($\varrho_0=0.16\,\mbox{fm}^{-3}$ being
the nuclear saturation density). The neutron pair wave function
demonstrates the BCS-BEC crossover behavior, being spatially
extended and of the oscillatory type
 at moderate densities and possessing the well
localized single peak at low densities.
 The spatial structure of the wave function of
two neutrons was studied also in the halo nucleus
$^{11}\mbox{Li}$~\cite{HSCS}. The behavior of the neutron pair at
different densities was simulated by calculating the pair wave
function at several distances between the core nucleus and the
center of mass of the two valence neutrons. Analogously to the
case of infinite nuclear matter, the same BCS- and BEC-like
structures in the pair wave function were found, corresponding to
the normal and low density regions, respectively.  In
Ref.~\cite{MSH}, the di-neutron correlations were studied on the
base of a new effective density dependent contact pairing
interaction. An important feature is that this contact interaction
contains the isospin dependence and was adjusted to reproduce the
pairing gap in symmetric nuclear matter and neutron matter,
obtained in the microscopic G-matrix calculations with the
realistic Argonne $v_{18}$ nucleon-nucleon (NN)
potential~\cite{CLS}. This study shows that for the screened
pairing interaction a di-neutron BEC state is formed in symmetric
nuclear matter at neutron densities
$\varrho_n/\varrho_0\sim10^{-3}$ and there is no evidence to the
appearance of the BEC state in neutron matter.

Since a two-neutron system in free space has a virtual state
around  zero energy, then the medium polarization effects should
play the major role in  the possible formation of the di-neutron
BEC state in low-density nuclear systems. However, the influence
of the nucleon medium on the di-neutron correlations seems to be
different in neutron matter and in symmetric nuclear matter. Most
of the recent quantum many-body calculations, e.g., based on the
Brueckner theory~\cite{CLS,LS}, or renormalization group
approach~\cite{SFB},  show the reduction of the pairing gap in
neutron matter by a factor of 2-3 due to the screening effect of
the pairing interaction. The only exception is the variational
quantum Monte Carlo calculations of Ref.~\cite{FFI}, showing a
small influence of the medium polarization on the neutron $^1S_0$
pairing. At the same time, calculations based on the Brueckner
theory show  that in symmetric nuclear matter the medium
polarization of the interaction enhances the di-neutron pairing
correlations due to  the proton particle-hole
excitations~\cite{CLS}. The same evidence in favor of  the medium
enhancement of the neutron $^1S_0$ pairing was obtained for finite
nuclei, where the pairing interaction becomes more attractive
because of  the surface vibrations~\cite{BBG,GBB,BBB}. Thus, one
can conclude that the conditions for the formation of a di-neutron
BEC state are more favorable in a nuclear environment rather than
in neutron matter.

The recent upsurge of interest to the BEC-BCS crossover is caused
by finding  the BCS pairing in ultracold  quantum atomic
gases~\cite{RGJ,ZSS,KHG}. Ultracold atomic gases provide an
experimental play-ground for testing pairing phenomena due to the
possibility to control the inter-atomic interactions via a
magnetically-tuned Feshbach resonance.   In fact, as one goes
through the resonance starting from the negative s-wave scattering
length and passing to its positive values, the Cooper pairs of two
atoms in two hyperfine states shrink in size and go over
continuously into the Bose-Einstein condensate of diatomic
molecules. In this respect, an important question arises  as to
what are the qualitative features allowing one to distinguish
between the BCS and BEC states.

In the low density limit, when  the interaction is characterized
by a single parameter, the s-wave scattering length $a$,  the
$^1S_0$ pairing can be described on the base of the so-called
regularized gap equation~\cite{ERS}. In this model, the strength
of the interaction is controlled only by the dimensionless
parameter $1/k_Fa$. The range $1/k_Fa\ll-1$  corresponds to the
weak coupling BCS regime while the range $1/k_Fa\gg1$ is related
to the strong coupling BEC regime.  The boundaries of the
crossover region between the BCS and BEC states, according to
Ref.~\cite{ERS}, are determined by the values $1/k_Fa=\pm1$. The
advantage of this regularized model is that all integrals in the
gap equation and the particle number equation can be expressed in
terms of some special functions and, hence, the model is easily
solvable. Then all quantities of interest can be calculated, in
particular, at the boundaries $1/k_Fa=\pm1$ of the crossover
region and these values can be used as the reference values for
the study of the BCS-BEC transition in a specific Fermi system at
low density.

Such a viewpoint was adopted in Refs.~\cite{M,MSH} for the study
of a possible BCS-BEC crossover of neutron pairs in nuclear
matter. %and its corresponding signatures.
 There were considered a few  probable characteristics of the crossover,
%considered in Refs.~\cite{M,MSH},
such as  the ratios $\xi_{\rm{rms}}/d_n,
\Delta_n/\varepsilon_{Fn}$~\cite{M,MSH},
$\mu_n/\varepsilon_{Fn}$~\cite{MSH}, and the probability $P(d_n)$
for the pair neutrons to come close to each other within the
relative distance $d_n$~\cite{M,MSH} ($\xi_{\rm{rms}},d_n,
\Delta_n,  \varepsilon_{Fn}$ and $\mu_n$ being the r.m.s. radius
of the Cooper pair, the mean average distance between neutrons,
the neutron energy gap,  the neutron Fermi kinetic energy, and the
effective neutron chemical potential, respectively). It was shown
that these criteria are closely related to each other and give
approximately the same density range $\varrho_n/\varrho_0\sim
10^{-4}-10^{-1}$ for the domain, where the neutron pairs are
strongly spatially correlated~\cite{M}. Besides, in terms of these
criteria
 the BCS-BEC
crossover was found for a di-neutron condensate in symmetric
nuclear matter with the density dependent contact interaction at
neutron densities $\varrho_n/\varrho_0\sim 10^{-3}$~\cite{MSH}.

However, it is necessary to note that in the case of nuclear
matter the assumption of the dilute gas limit may be justified
only at very low densities $\varrho_n/\varrho_0\lesssim
10^{-5}$~\cite{LS}. Hence, the use of the regularized model with
the contact interaction for getting the reference values of
various criteria   can lead only to the qualitative results at the
above densities.

   Thus, it is desirable to approach the problem from the different point of
view and to look for  some other possible characteristics for
identifying the BCS and BEC states. In this respect, using the
analogies with other low-density Fermi systems, such as, e.g.,
ultracold atomic gases, could be expedient. To that end, let us
mention the idea to utilize the density-density correlations in
the image of an expanding gas cloud to detect superfluid states in
the system of fermionic atoms released from the trap~\cite{ADL}.
This idea was explored in Ref.~\cite{MGB} to study the alterations
in the density-density correlations through the BCS-BEC crossover.
It was learned  that the density correlation function of
two-component ultracold fermionic atomic gas with singlet pairing
of fermions changes sign at low momentum transfer across the
BCS-BEC transition. This feature was considered as an unambiguous
signature of the crossover. Here we would like to check this
criterion with respect to  the BCS-BEC crossover in a di-neutron
condensate of nuclear matter. Besides, the calculations of
Ref.~\cite{MGB}, performed for the balanced populations of
fermions of two species,  will be extended to unbalanced
populations of unlike fermions (for neutron-proton systems, this
means finite isospin asymmetry). In defining a primary
characteristic for identifying  the BCS-BEC transition, we adopt a
viewpoint that the qualitative boundary between the BCS and BEC
states occurs when the chemical potential reaches the zero value.
Just this feature was considered as an indication of the BCS-BEC
crossover in the earlier researches  on the BCS-BEC transition
 in nuclear matter~\cite{BLS,AFRS}.

\section{Basic equations}
 The  equation for the neutron energy gap in
the $^1S_0$ pairing channel reads~\cite{AIP} \beqe \Delta_n({\bf
k}) =-\frac{1}{V}\sum_{{\bf k}'}V({\bf k},{\bf
k}')\frac{\Delta_n({\bf k}')}{2E_n({\bf k}')}\tanh\frac{E_n({\bf
k}')}{2T}. \label{Gap}\eeqe Here
$$E_n({\bf k})=\sqrt{\xi_n^2({\bf k})+\Delta_n^2({\bf k})},\,
\xi_n({\bf k})=\varepsilon_n({\bf k})-\mu_{0n},$$
$\varepsilon_n({\bf k})$ and $\mu_{0n}$ being  the neutron  single
particle energy and neutron chemical potential,  respectively.
Further the single particle energy will be taken in the form
corresponding to the Skyrme effective interaction,
$\varepsilon_n=\hbar^2k^2/2m_n+U_n$, with $m_n$ and $U_n$ being
the neutron effective mass and neutron single particle potential,
respectively.  The quadratic on momentum term  in the neutron
single particle potential is already included into the kinetic
energy term and the remaining part $U_n$ is momentum independent
and can be incorporated into the effective neutron
 chemical potential $\mu_n\equiv\mu_{0n}-U_n$.  Eq.~\p{Gap} should be
solved self-consistently with the equation for the neutron
particle number density $\varrho_n$,\beqe
\varrho_n=\frac{2}{V}\sum_{\bf
k}\frac{1}{2}\Bigl(1-\frac{\xi_n({\bf k})}{E_n({\bf
k})}\tanh\frac{E_n({\bf k})}{2T} \Bigr)\equiv\frac{2}{V}\sum_{\bf
k}f_{\bf k}.\label{Dens}\eeqe

Let us introduce the neutron anomalous distribution
function~\cite{AIP}
$$\psi({\bf k})=\langle a^+_{n,{\bf k\uparrow}}a^+_{n,-{\bf k\downarrow}}
\rangle=\frac{\Delta_n({\bf k})}{2E_n({\bf k})}\tanh\frac{E_n({\bf
k})}{2T}, $$ where $\langle...\rangle\equiv \mbox{Tr} \varrho...$,
$\varrho$ being the density matrix of the system. Then, using
Eq.~\p{Dens}, one can represent Eq.~\p{Gap} for the energy gap in
the form \beqe\frac{\hbar^2k^2}{m_n}\psi({\bf k})+(1-2f_{\bf
k})\sum_{{\bf k}'}V({\bf k},{\bf k}')\psi({\bf
k}')=2\mu_n\psi({\bf k}).\label{Schr}\end{equation} In the limit
of vanishing density, $f_k\rightarrow0$, Eq.~\p{Schr} goes over
into the Schr\"odinger equation for a di-neutron bound state. The
corresponding energy eigenvalue is equal to $2\mu_n$. The change
of  sign of the effective neutron chemical potential $\mu_n$
  under decreasing density of nuclear matter signals the
transition from the regime of large overlapping neutron Cooper
pairs to the regime of non-overlapping di-neutron bound states.

Let us consider the two-neutron density correlation function \bal
{\tcal D} (\tbf {x},\tbf {x}')&=\langle\Delta\hat
n(\tbf{x})\Delta\hat n(\tbf
{x}')\rangle,\; \Delta\hat n(\tbf{x})=\hat n(\tbf{x})-\hat n,\label{12n}\\
\hat n(\tbf{x})& =
\frac{1}{V}\sum_{\sigma\tbf{kk}'}e^{i(\tbf{k}'-\tbf{k})\tbf{x}}a^+_{n,\tbf{k}\sigma}
a_{n,\tbf{k}'\sigma},\;\nonumber\\
\hat n&= \frac{1}{V}\sum_{\sigma\tbf{k}}a^+_{n,\tbf{k}\sigma}
a_{n,\tbf{k}\sigma}.\nonumber \end{align} Its general structure in
the spatially uniform and isotropic case reads~\cite{LL5} \beq
{\tcal D}(\tbf {x},\tbf {x}')=\varrho_n\delta(\tbf{r})+\varrho_n
D(r),\; \tbf{r}=\tbf {x}-\tbf {x}'.\end{eqnarray} The function
$D(r)$ is called the density correlation function as well. We will
just be  interested in the behavior of the function $D(r)$.
Calculating the trace in Eq.~\p{12n}  and going to the Fourier
representation
$$D(q)=\int d\tbf {r}e^{i\tbf {q}\tbf
{r}}D(r),$$ one can get \beq D(q)=I_\psi(q)-
I_f(q),\label{D}\end{eqnarray} where \bal
I_f(q)&=\frac{2}{\pi^3\varrho_n}\int_0^\infty
dr\,r^2j_0(rq)\Bigl[\int_0^\infty
dk\,k^2f(k)j_0(rk)\Bigr]^2,\nonumber\\
I_\psi(q)&=\frac{2}{\pi^3\varrho_n}\int_0^\infty
dr\,r^2j_0(rq)\Bigl[\int_0^\infty
dk\,k^2\psi(k)j_0(rk)\Bigr]^2.\nonumber \end{align} Here $j_0$ is
the spherical Bessel function of the first kind of  order zero.
The functions $I_f$ and $I_\psi$ represent the normal and
anomalous contributions to the neutron density correlation
function.

Analogously, we can consider the two-neutron spin correlation
function \bal{{\tcal S}}_{\mu\nu} (\tbf {x},\tbf
{x}')&=\langle\Delta\hat s_\mu(\tbf{x})\Delta\hat s_\nu(\tbf
{x}')\rangle,\; \Delta\hat s_\mu(\tbf{x})=\hat s_\mu(\tbf{x})-\hat s_\mu,\label{13}\\
\hat s_\mu(\tbf{x}) &=\frac{1}{2V}\sum_{\sigma\sigma '  \tbf{kk}'}
e^{i(\tbf{k}'-\tbf{k})\tbf{x}}a^+_{n,\tbf{k}\sigma}(\sigma_\mu)_{\sigma\sigma'}
a_{n,\tbf{k}'\sigma'},\,\nonumber\\
\hat s_\mu&=
\frac{1}{2V}\sum_{\sigma\sigma'\tbf{k}}a^+_{n,\tbf{k}\sigma}(\sigma_\mu)_{\sigma\sigma'}
a_{n,\tbf{k}\sigma'},\nonumber
\end{align}where $\sigma_\mu$ are the Pauli matrices.
The general structure of the neutron spin correlation function for
the spin unpolarized case is \beqe {\cal
S}_{\mu\nu}(\tbf{x,x}')=\frac{\varrho_n}{4}\,\delta_{\mu\nu}\,\delta(\tbf{r})+\varrho_n\,
S_{\mu\nu}(r).\label{15}\eeqe Calculating the trace in Eq.~\p{13},
for the Fourier transform of the spin  correlation function one
gets   \beqe S_{\mu\nu}(q)=S(q)\delta_{\mu\nu},\; S(q)=
-\frac{1}{4}(I_f(q) +I_\psi(q)).\label{S}\eeqe

At zero momentum transfer, the normal $I_f$ and anomalous $I_\psi$
contributions read \bal
 I_f(q=0) &=\frac{1}{\pi^2\varrho_n}\int_0^\infty
dk\,k^2f^2({k}),\label{I_0}\\I_\psi(q=0)
&=\frac{1}{\pi^2\varrho_n}\int_0^\infty dk\,k^2\psi^2({k}).
\nonumber\end{align}

At zero temperature, the zero-momentum transfer neutron spin
correlation function satisfies the sum rule \beqe S(q=0)
=-\frac{1}{4}(I_f(0) +I_\psi(0))=
-\frac{1}{4}\,.\label{sum_rule}\eeqe

\section{$^1S_0$ neutron energy gap at zero temperature}

Further we will take the pairing interaction in the $^1S_0$
neutron pairing channel in the form of the effective density
dependent contact interaction
\begin{equation} V({\bf r}_1,{\bf r}_2)=v_0g(\varrho_n,\varrho_p)\delta({\bf r}_1-{\bf
r}_2)\label{V}.\end{equation} In the gap equation, the
interaction~\p{V} should be supplemented with the cut-off energy
$E_c$. The interaction strength $v_0$ is determined in such a way
that the interaction~\p{V} reproduces the low-energy
neutron-neutron scattering phase shifts obtained with the Argonne
$v_{18}$ potential. The isospin dependent factor $g$ was
constructed in Ref.~\cite{MSH}  and reads
\bal g&=g_1+g_2,\label{g}\\
g_1&=1-(1-\alpha)\eta_s\biggl(\frac{\varrho}{\varrho_0}\biggr)^{\beta_s}-
\alpha\eta_n\biggl(\frac{\varrho}{\varrho_0}\biggr)^{\beta_n},\nonumber\\
g_2&=\eta_2\biggl[\biggl(1+e^\frac {k_{Fn}-1.4}{0.05}\biggr)^{-1}-
\biggl(1+e^\frac{
k_{Fn}-0.1}{0.05}\biggr)^{-1}\biggr].\nonumber\end{align} Here
$\alpha\equiv(\varrho_n-\varrho_p)/\varrho$ is the isospin
asymmetry parameter, $k_{Fn}\equiv(3\pi^2\varrho_n)^{1/3}$ is the
neutron Fermi momentum, $\beta_s,\beta_n,\eta_s,\eta_n,\eta_2$ are
the fitting parameters, adjusted to reproduce the pairing gap in
symmetric nuclear matter and neutron matter obtained in the
G-matrix calculations with the Argonne $v_{18}$ potential with
account of the medium polarization effects~\cite{CLS}. The term
$g_1$ contains the density dependence  in a power law form as it
was introduced  in the contact interaction  of Ref.~\cite{BE}.
Besides, the term $g_1$ contains the isospin dependence, built
such as to reproduce the position and the magnitude of the maximum
of the energy gap in both symmetric nuclear matter and neutron
matter. The term $g_2$
 is necessary to improve the results of the fitting
procedure for the energy gap at the neutron Fermi momenta
$k_{Fn}>1\,\mbox{fm}^{-1}$. Concerning the term $g_2$, let us
make one more comment. In the original article~\cite{MSH}, this
term is written as $g_2=g_2(k_F)$, where the quantity $k_F$ is not
explicitly determined. There are two possibilities: (1)
$g_2=g_2(k_F)$, where $k_F$ is parametrized in terms  of the total
density $k_F=(3\pi^2\varrho/2)^\frac{1}{3}$, and (2)
$g_2=g_2(k_{Fn})$ with $k_{Fn}$ being the neutron Fermi momentum.
 There will be no difference between these two
parametrizations for symmetric nuclear matter; however, the
difference occurs for isospin asymmetric matter and will be most
serious for neutron matter. The calculations of the energy gap in
neutron matter show that   the results of the fitting procedure,
presented in Ref.~\cite{MSH}, correspond to the case
$g_2=g_2(k_{Fn})$. This point was confirmed in the
communication~\cite{JM}. Therefore, the term $g_2$ is, in fact,
isovector and not isoscalar, as it is called in Ref.~\cite{MSH}.
Since the  microscopically  calculated  neutron energy gap in
symmetric nuclear matter and neutron matter is well fitted with
$g_2=g_2(k_{Fn})$, one can use this form also for calculations at
intermediate asymmetries $0<\alpha<1$. Besides, if to apply the
isospin dependent contact interaction~\p{V}, \p{g} for the
calculation of the pairing gaps in finite nuclei~\cite{MSH2}, one
has to change $\alpha\rightarrow-\alpha$ in the term $g_1$ and to
use $g_2=g_2(k_{Fp})$ ($k_{Fp}$ being the proton Fermi momentum)
in the proton $^1S_0$ pairing channel.

\begin{table}[tb]
\caption{\label{tab1}The parameters of the screened-II
interaction~\cite{MSH} obtained from the fit to the neutron energy
gap in symmetric nuclear matter and neutron matter, calculated for
the Argonne $v_{18}$ potential with  account of the medium
polarization effects.}
\begin{ruledtabular}
\begin{tabular}{ccccccc}
 $v_0$ (MeV\,fm$^3$) & $E_c$ (MeV) & $\eta_2$ & $\eta_s$ & $\beta_s$ & $\eta_n$ & $\beta_n$ \\\hline
   -542 & 40 & 0.8 & 1.80 & 0.27 & 1.61 & 0.122\\
\end{tabular}
\end{ruledtabular}
\end{table}
%\end{verbatim}

 Let us now turn to the calculation of the neutron energy gap in the $^1S_0$
pairing channel that is the necessary preliminary step before the
calculation of the correlation functions. The zero temperature gap
equation  with the effective contact interaction~\p{V} reads \beqe
\Delta_n=-v_0g\int_{k_-<k<k_+}\
\frac{d^3k}{(2\pi)^3}\frac{\Delta_n}{2E_n(k)}. \label{gap} \eeqe

 In the gap equation,  the cut-off momenta $k_\pm$ are
defined from the requirement that the quasiparticle energy should
be less than the cut-off energy $E_c$, $E_n(k)<E_c$. If $k_-$
becomes imaginary,   one should set $k_-\equiv0$. Just this choice
of the cut-off parameters is often used in the HFB calculations.
Eq.~\p{gap} should  be solved self-consistently with Eq.~\p{Dens}
for the neutron particle number density. In numerical
calculations, we  use the screened-II interaction of
Ref.~\cite{MSH} with the parameters given in Table 1. The single
particle energy $\varepsilon_n(k)$ is taken  for the SLy4 Skyrme
effective interaction.

%\begin{verbatim}
\begin{figure}[tb] % fig 1
\begin{center}
\includegraphics[height=12.6cm,width=8.6cm,trim=49mm 102mm 54mm 46mm,
draft=false,clip]{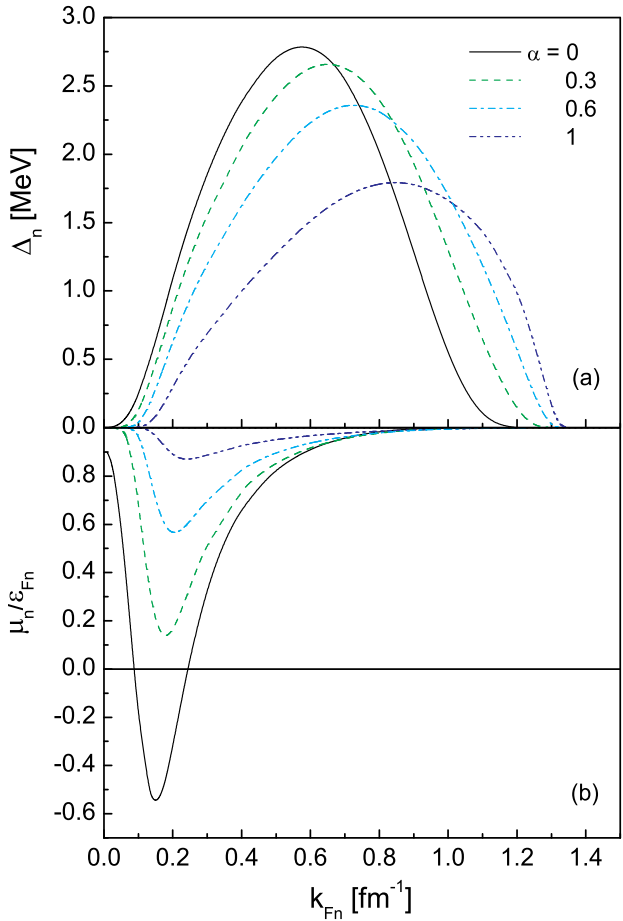}
\end{center}
\caption{(Color online) (a) The zero temperature neutron energy
gap and (b) the ratio $\mu_n/\varepsilon_n$ of the effective
neutron chemical potential to the neutron Fermi kinetic energy
as functions of the neutron Fermi momentum at different isospin
asymmetries for the screened-II contact interaction.} \label{fig1}
\end{figure}

In Fig.~\ref{fig1}a, the neutron energy gap is plotted as a
function of the neutron Fermi momentum at zero temperature and
different isospin asymmetries. With increasing asymmetry, the
magnitude of the energy gap decreases and the position of its
maximum  is shifted to the larger neutron Fermi momentum.   In
Fig.~\ref{fig1}b, the ratio of the effective neutron chemical
potential $\mu_n$ to the neutron Fermi kinetic energy
$\varepsilon_{Fn}=\hbar^2k^2_{Fn}/2m_n$ is depicted as a function
of the neutron Fermi momentum. This ratio is indicative of the
importance of the di-neutron correlations, since in the absence of
pairing correlations it is equal to unity, otherwise it is less
than unity. It is seen that the di-neutron correlations are most
strong in symmetric nuclear matter and with increasing isospin
asymmetry their strength is diminished. In accordance with
Eq.~\p{Schr},  we will identify the BCS and BEC states of the pair
condensate by the positive (BCS regime) and negative (BEC regime)
values of the effective chemical potential $\mu_n$. For symmetric
nuclear matter, a di-neutron BEC state is realized at the finite
density range $0.09\,\mbox{fm}^{-1} < k_{Fn}<
0.25\,\mbox{fm}^{-1}$ and surrounded on both sides by the BCS
state. For neutron matter, the self-consistent equations have no
at all solutions with the negative chemical potential $\mu_n$ and
the BEC state is not formed. For asymmetric nuclear matter, the
BEC state is quite sensitive to the isospin asymmetry and is
quenched at the critical value  $\alpha_c\approx 0.26$. Note that,
as was clarified in Ref.~\cite{CLS}, the medium polarization acts
differently on the di-neutron correlations at low-density region,
increasing the neutron energy gap in symmetric nuclear matter and
decreasing it in neutron matter.

\begin{figure}[tb] % fig 1
\begin{center}
\includegraphics[height=12.6cm,width=8.6cm,trim=49mm 102mm 54mm 46mm,
draft=false,clip]{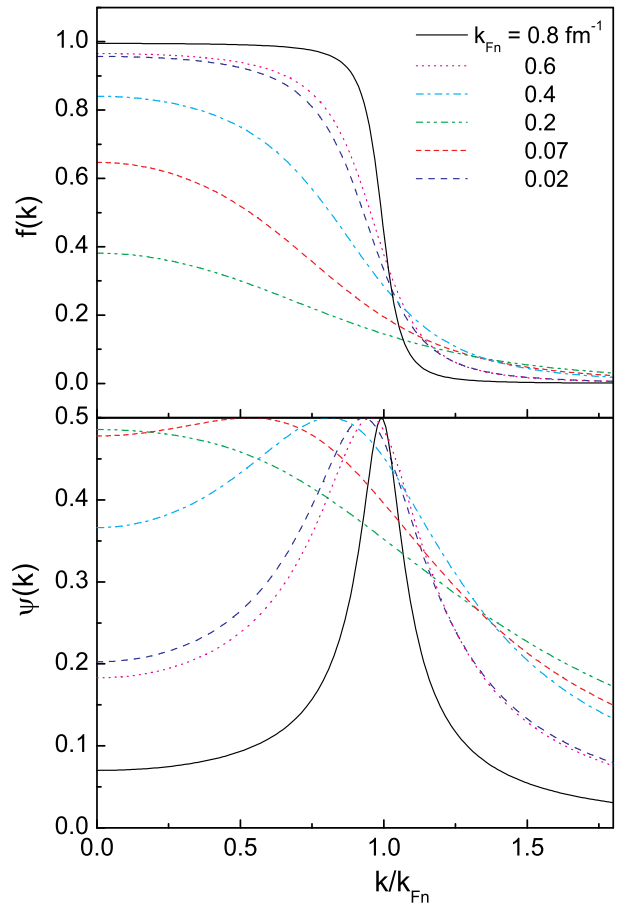}
\end{center}
\caption{(Color online)  The normal  $f(k)$ and   anomalous
$\psi(k)$ distribution functions  of neutrons in symmetric nuclear
matter  at different neutron Fermi momenta.} \label{fig2}
\end{figure}

Fig.~\ref{fig2} shows the evolution of the neutron normal $f(k)$
and anomalous $\psi(k)$ distribution functions with density in
symmetric nuclear matter. In the weak coupling BCS regime
($k_{Fn}=0.8\,\mbox{fm}^{-1}$), the normal distribution function
resembles the Fermi step distribution function with the edge near
the neutron Fermi momentum. With decreasing density, the step-like
structure is gradually washed out and in the region of the  strong
coupling BEC regime ($k_{Fn}=0.2\,\mbox{fm}^{-1}$) the normal
distribution function  smoothly changes with  the momentum.
However, under further decreasing density, we go back to the BCS
regime ($k_{Fn}=0.02\,\mbox{fm}^{-1}$) and the step-like structure
becomes more and more pronounced. The anomalous distribution
function $\psi(k)$ in the BCS regime has  a well developed peak,
indicating that pairing exists only near the Fermi surface. In the
strong coupling BEC regime $\psi(k)$ exhibits a smooth variation
typical of the wave function of a bound state. At ultra-low
densities, the spike in the momentum dependence of  the anomalous
distribution function appears again.

\begin{figure}[tb] % fig 1
\begin{center}
\includegraphics[height=12.6cm,width=8.6cm,trim=47mm 102mm 54mm 46mm,
draft=false,clip]{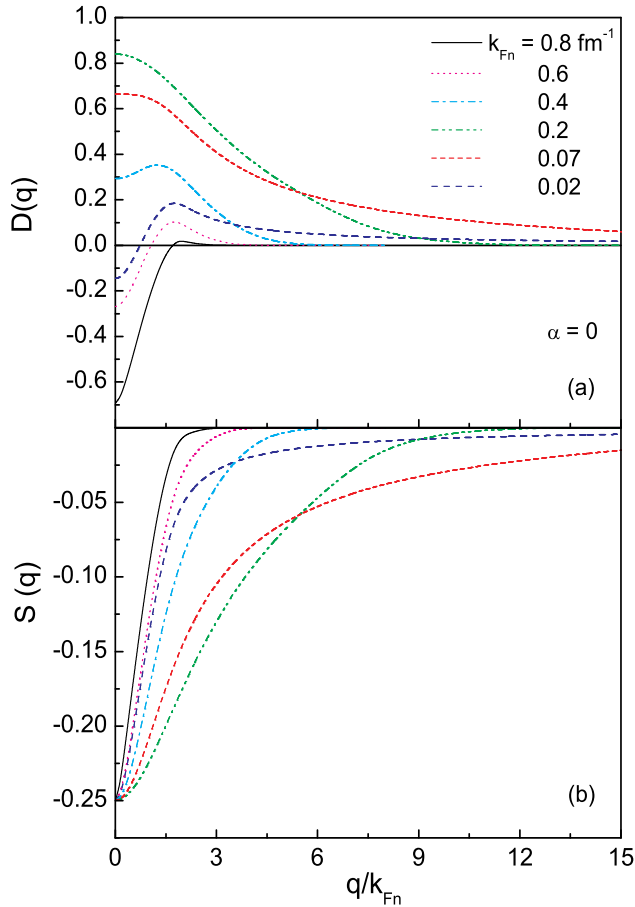}
\end{center}
\caption{(Color online) (a) Density $D(q)$ and (b) spin $S(q)$
correlation functions as functions of the momentum transfer at
different neutron Fermi momenta in symmetric nuclear matter.}
\label{fig3}
\end{figure}

\begin{figure}[tb] % fig 1
\begin{center}
\includegraphics[height=12.6cm,width=8.6cm,trim=47mm 102mm 54mm 46mm,
draft=false,clip]{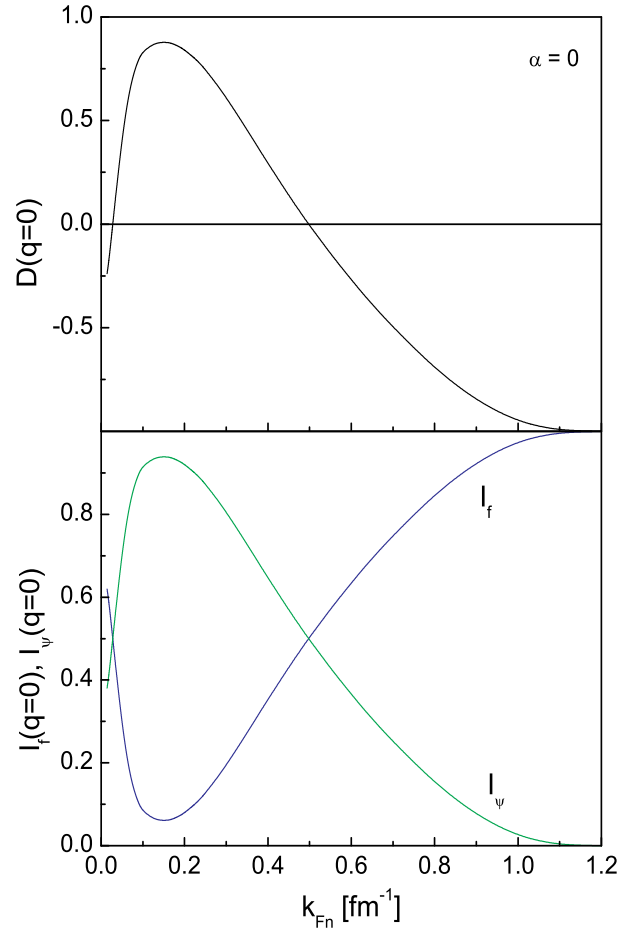}
\end{center}
\caption{(Color online) Zero-momentum transfer density correlation
function $D(q=0)$ and its normal $I_f(q=0)$ and anomalous
$I_\psi(q=0)$ contributions  as functions of the neutron Fermi
momentum in symmetric nuclear matter.} \label{fig4}
\end{figure}

\section{Density and spin correlation functions for a di-neutron
condensate at zero temperature}
\subsection{Symmetric nuclear matter}
 After
the neutron energy gap and the effective neutron  chemical
potential have been found, the density $D(q)$ and spin $S(q)$
correlation functions can be obtained directly from Eqs.~\p{D} and
\p{S}. Fig.~\ref{fig3}a shows the density correlation function as
a function of the momentum transfer for a set of the neutron Fermi
momenta in symmetric nuclear matter. Let us begin with the
consideration of the BCS regime at the moderate densities
($k_{Fn}=0.8\,\mbox{fm}^{-1}$), where at low momentum transfer
$D(q)$ is negative. Under decreasing density ($k_{Fn}=0.6,
0.4\,\mbox{fm}^{-1}$), the zero-momentum transfer density
correlation function  $D(q=0)$ at certain neutron Fermi momentum
changes  sign
 from negative  to  positive. In the BEC regime ($k_{Fn}=0.2\,\mbox{fm}^{-1}$),
the correlation function $D(q)$ is everywhere
 positive. If to decrease density further ($k_{Fn}=0.07, 0.02\,\mbox{fm}^{-1}$),
 the zero-momentum transfer
value $D(q=0)$ changes sign again, now from  positive  to
negative, and in the ultra-low density BCS regime
 the function $D(q)$ has the negative
sign at low momentum transfer as it was  at the moderate
densities. The change of  sign of the density correlation function
at low momentum transfer across the BCS-BEC transition region was
considered in Ref.~\cite{MGB} as a signature of the crossover. We
see that this feature is qualitatively well reproduced for a
di-neutron condensate in symmetric nuclear matter. At the same
time, the spin correlation function, as can be seen from
Fig.~\ref{fig3}b, fluently evolves across the BCS-BEC transition
region without any qualitative change.
\begin{figure}[tb] % fig 1
\begin{center}
\includegraphics[height=10.5cm,width=8.6cm,trim=53mm 124mm 54mm 47mm,
draft=false,clip]{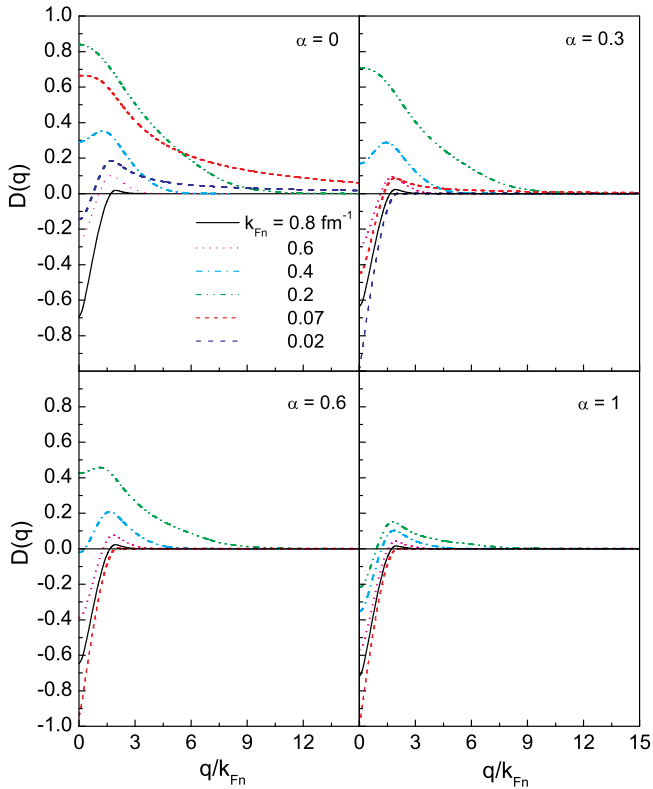}
\end{center}
\caption{(Color online) Density correlation function $D(q)$ as a
function of the momentum at different neutron Fermi momenta and
isospin asymmetries.} \label{fig5}
\end{figure}
In order to establish the boundaries  of the density interval for
the BEC state  on the basis of the formulated criterion, let us
consider the density correlation function at zero momentum
transfer, depicted in Fig.~\ref{fig4}. Under decreasing density,
the density correlation function $D(q=0)$ twice changes sign,
first, at the point of the  BCS-to-BEC transition and then at the
point of the reverse BEC-to-BCS crossover, that corresponds to the
earlier consideration. The BEC state is realized in the range
$0.03\,\mbox{fm}^{-1} < k_{Fn}< 0.5\,\mbox{fm}^{-1}$ and this
interval is wider than that determined on the basis of the
negative values of the effective neutron chemical potential.

The normal $I_f(q=0)$ and anomalous $I_\psi(q=0)$ contributions to
$D(q=0)$, given by Eq.~\p{I_0},  are shown in the bottom panel of
Fig.~\ref{fig4}. This figure qualitatively explains  why the
density correlation function $D(q=0)$   changes  sign at the
boundary between the  BCS and BEC states. In the weak coupling BCS
regime ($k_{Fn}=0.8\,\mbox{fm}^{-1}$),   the anomalous
distribution function $\psi(k)$ is sharply peaked near $k=k_{Fn}$,
as shown in Fig~\ref{fig2}, and, hence, the anomalous contribution
to $D(q=0)$ is much smaller than the normal one. Therefore,
$D(q=0)<0$ on the   BCS side from the transition boundary.
However, in the strong coupling BEC regime
($k_{Fn}=0.2\,\mbox{fm}^{-1}$), the normal distribution function
$f(k)$ is to a considerable extent depleted and the anomalous
contribution dominates over the normal one. Hence, $D(q=0)>0$ on
the BEC side from the transition boundary,  and at some critical
density between the BCS and BEC regimes the density correlation
function $D(q=0)$ should change the sign. Note also that,
according
  to the sum rule in Eq.~\p{sum_rule}, the normal and anomalous contributions satisfy
the relationship $I_f(0)+I_\psi(0)=1$.

As  seen from the figure, the contribution of the anomalous part
is the largest at the neutron Fermi momentum $k_{Fn}\sim
0.2\,\mbox{fm}^{-1}$. At the corresponding densities
$\varrho_n/\varrho_0\sim 10^{-3}$ the BEC di-neutron state can be
considered as well formed. Just at these densities the BEC of
bound neutron pairs in symmetric nuclear matter was found in
Ref.~\cite{MSH} with the help of other quantitative
characteristics, mentioned in Introduction.

\subsection{Asymmetric nuclear matter}
Now we will present the results of the numerical determination of
the correlation functions at finite isospin asymmetry. The
evolution of the density correlation function $D(q)$ with the
isospin asymmetry is shown in Fig.~\ref{fig5}. If to identify the
  BCS-BEC transition by the change of  sign of
the density correlation function at low momentum transfer, it is
seen  that with increasing asymmetry the density interval, where
the BEC state exists, is contracted.  At strong enough isospin
asymmetry only the BCS state is realized, as can be seen from the
limiting case $\alpha=1$, corresponding to neutron matter.

\begin{figure}[t] % fig 1
\begin{center}
\includegraphics[height=7.0cm,width=8.6cm,trim=48mm 142mm 56mm 66mm,
draft=false,clip]{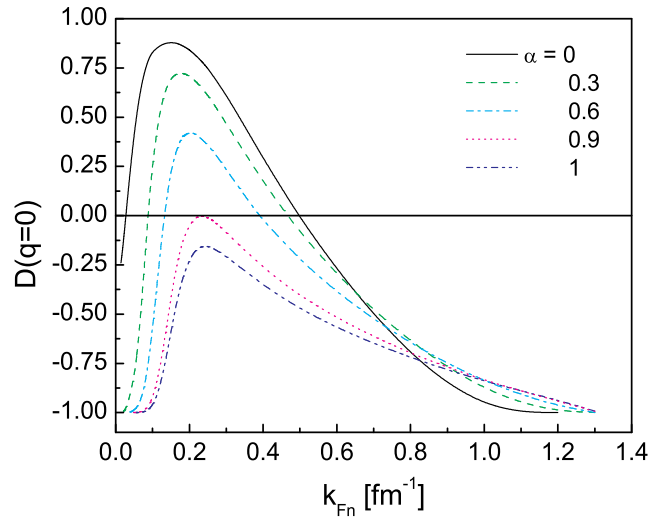}\end{center} \caption{(Color online)
Zero-momentum transfer density correlation function $D(q=0)$   as
a function of the neutron Fermi momentum at different isospin
asymmetries.} \label{fig6}
\end{figure}
In order to find an estimate for the critical isospin asymmetry,
at which the BEC state disappears, let us consider the
zero-momentum transfer density correlation function $D(q=0)$ as a
function of the neutron Fermi momentum   at different asymmetries
(Fig.~\ref{fig6}). For large isospin asymmetry
$\alpha>\alpha_c^d\sim 0.9$ the function $D(q=0)$ is always
negative that corresponds to the BCS state for all densities where
a di-neutron condensate exists. This critical asymmetry is much
larger than $\alpha_c\approx 0.26$, found  earlier on the basis of
the change of  sign of the chemical potential $\mu_n$. From that
one can infer that the criterion of the BCS-BEC crossover, based
on the change of  sign of the density correlation function, can be
trusted only at small isospin asymmetry. This conclusion
qualitatively agrees with that of Ref.~\cite{I}, where the density
and spin correlation functions were calculated for  a
neutron-proton condensate in dilute asymmetric nuclear matter.

\begin{figure}[tb] % fig 1
\begin{center}
\includegraphics[height=10.5cm,width=8.6cm,trim=50mm 124mm 54mm 47mm,
draft=false,clip]{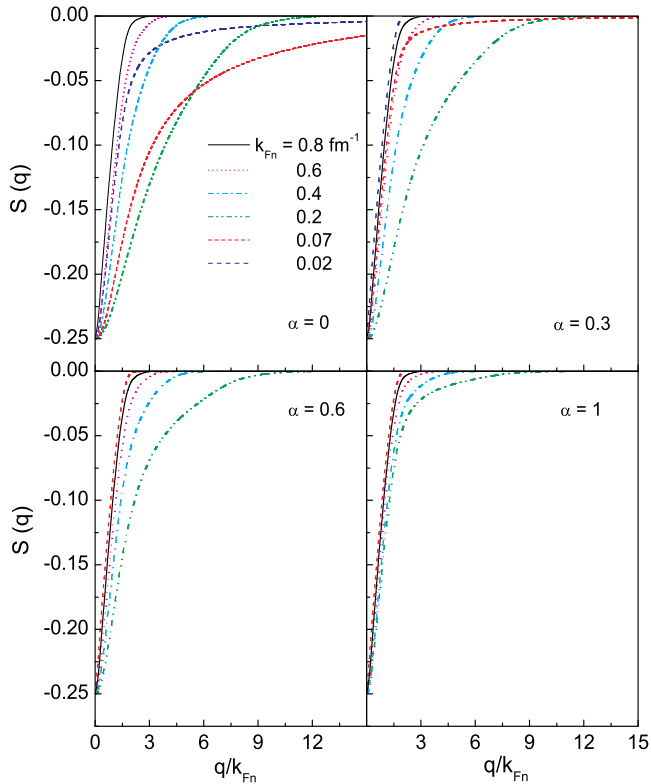}
\end{center}
\caption{(Color online) Spin correlation function $S(q)$ as a
function of the momentum at different neutron Fermi momenta and
isospin asymmetries.} \label{fig7}
\end{figure}

Fig.~\ref{fig7} shows the evolution of the spin correlation
function $S(q)$ with the isospin asymmetry. It is seen that the
zero-momentum transfer value $S(q=0)$ satisfies the sum
rule~\p{sum_rule}, independently of the isospin asymmetry. The
spin correlation function $S(q)$ smoothly varies between its
limiting values without any qualitative change.

In summary, the density and spin correlation functions have been
calculated for a di-neutron condensate in asymmetric nuclear
matter with the aim of finding the possible signatures of the
BCS-BEC crossover. As the primary characteristic for identifying
the transition, it is accepted that the qualitative boundary
between the BCS and BEC states occurs when the  effective neutron
chemical potential reaches the zero value. It has been shown that
the zero-momentum transfer spin correlation function satisfies the
sum rule at zero temperature. In symmetric nuclear matter, the
density correlation function changes sign at low momentum transfer
across the BCS-BEC transition  and this feature can be considered
as a signature of the crossover. This result qualitatively  agrees
with that obtained for two-component ultracold fermionic atomic
gas with equal densities of fermions of different
species~\cite{MGB}.
 The self-consistent calculations of the
$^1S_0$ energy gap in a di-neutron condensate and the effective
neutron chemical potential $\mu_n$ show  that the BEC state
($\mu_n<0$) is quite sensitive to isospin asymmetry and does not
appear at $\alpha>\alpha_c\approx0.26$. This is a reflection of
the fact that the medium polarization effects act differently on
di-neutron correlations at small and strong isospin asymmetry,
enhancing them in symmetric nuclear matter and suppressing them in
neutron matter. At finite isospin asymmetry, the criterion, based
on the change of  sign of the density correlation function,  gives
too large value for the critical asymmetry $\alpha_c^d\sim0.9$, at
which the BEC state is quenched. Therefore, it can be trusted for
the description of the density-driven BCS-BEC crossover of neutron
pairs  only at small isospin asymmetry. This result generalizes
the conclusion of Ref.~\cite{MGB}, in which the change of sign of
the density correlation function at low momentum transfer in
two-component quantum fermionic atomic gas with the balanced
populations of fermions of different species was considered as an
unambiguous signature of the BCS-BEC transition.

%\newpage


\begin{thebibliography}{10}
\bibitem{GSSN} G. R\"opke, A. Schnell, P. Schuck, and P. Nozieres,
Phys.  Rev.  Lett.  {\bf 80}, 3177 (1998).
\bibitem{BLS} M. Baldo, U. Lombardo and P. Schuck,  Phys. Rev. C {\bf 52},
975 (1995).
\bibitem{BSKR}  M. Beyer, W. Schadow, C. Kuhrts, and G. R\"opke, Phys. Rev. C {\bf 60},
034004 (1999).
\bibitem{SC} A. Sedrakian, and J.W. Clark,
  Phys. Rev. C {\bf 73}, 035803  (2006).
 \bibitem{E} D.M. Eagles,  Phys.  Rev.   {\bf 186}, 456 (1969).
\bibitem{L} A.J. Leggett, J. Phys. C (Paris)
{\bf 41}, 7 (1980).
\bibitem{NS} P. Nozieres and S. Schmitt--Rink,   J. Low Temp. Phys. {\bf 59},
195 (1985).
\bibitem{AFRS} T. Alm, B. L. Friman, G. R\"opke, and H. Schulz,   Nucl. Phys. {\bf A551},
45 (1993).
\bibitem{SL} A.
Sedrakian and U.  Lombardo,  Phys.  Rev.  Lett.  {\bf 84}, 602
(2000).
 \bibitem{AIPY} A.I.  Akhiezer, A.A.  Isayev, S.V.  Peletminsky, and
 A.A. Yatsenko,    Phys.  Rev.  C {\bf 63}, 021304(R) (2001).
 \bibitem{LNS} U. Lombardo, P. Nozieres, P. Schuck, H.-J. Schulze, and
 A. Sedrakian,
  Phys. Rev. C {\bf 64}, 064314 (2001).
 \bibitem{IYB} A.A.  Isayev, S.I.  Bastrukov,
  and J. Yang,   Nuclear Physics    {\bf A734}, E112 (2004);
  Physics of Atomic Nuclei { \bf 67}, 1840 (2004).
\bibitem{M}M. Matsuo, Phys. Rev. C {\bf 73}, 044309 (2006).
\bibitem{HSCS} K. Hagino, H. Sagawa, J. Carbonell, and P. Schuck, Phys. Rev.
Lett. {\bf 99}, 022506 (2007).
\bibitem{MSH}J. Margueron, H. Sagawa, and K. Hagino,  Phys. Rev. C {\bf 76}, 064316 (2007).
\bibitem{CLS}L. G. Cao, U. Lombardo, and P. Schuck, Phys. Rev. C {\bf 74}, 064301
(2006).
\bibitem{LS} U. Lombardo and H.-J. Schulze, "Superfluidity in
Neutron Star Matter" in {\it Physics of Neutron Star Interiors},
Lecture Notes in Physics, Vol. 578, pp. 30-54. Eds. D. Blaschke,
N.K. Glendenning and A. Sedrakian (Springer Verlag, 2001).
\bibitem{SFB}
A. Schwenk, B. Friman, and G. E. Brown, Nucl. Phys. {\bf A713},
191 (2003).
\bibitem{FFI}
A. Fabrocini, S. Fantoni, A. Yu. Illarionov, and K. E. Schmidt,
Phys. Rev. Lett. {\bf 95}, 192501 (2005).
\bibitem{BBG}
F. Barranco, R. A. Broglia, G. Gori, E. Vigezzi, P.-F. Bortignon,
and J. Terasaki, Phys. Rev. Lett. {\bf 83}, 2147 (1999).
\bibitem{GBB}
N. Giovanardi, F. Barranco, R. A. Broglia, and E. Vigezzi, Phys.
Rev. C {\bf 65}, 041304(R) (2002).
\bibitem{BBB}
F. Barranco, P. F. Bortignon, R. A. Broglia, G. Colo, P. Schuck,
E. Vigezzi, and X. Vi$\tilde{\mbox n}$as, Phys. Rev. C {\bf 72},
054314 (2005).
\bibitem{RGJ}  C.A. Regal, M. Greiner, and D.S. Jin,  Phys.  Rev.
Lett.
 {\bf 92}, 040403 (2004).
\bibitem{KHG}  J. Kinast, S.L. Hemmer, M.E. Gehm, A. Turlapov, and J.E. Thomas,
  Phys.  Rev. Lett.
 {\bf 92}, 150402 (2004).
\bibitem{ZSS} M.W. Zwierlein, C.A. Stan, C.H. Schunck, S.M.F. Raupach, A.J. Kerman, and W. Ketterle,
 Phys.  Rev. Lett. {\bf 92}, 120403 (2004).
\bibitem{ERS} J. R. Engelbrecht, M. Randeria, and C. A. R. S\'a de Melo, Phys.
Rev. B {\bf 55}, 15153 (1997).
\bibitem{ADL} E. Altman, E. Demler, and M. D. Lukin, Phys.
Rev. A {\bf 70}, 013603 (2004).
 \bibitem{MGB} B. Mihaila, S. Gaudio, K.B. Blagoev, A.V. Balatsky, P.B. Littlewood,
 and D.L. Smith, Phys. Rev. Lett.
{\bf 95}, 090402 (2005).
\bibitem{AIP} A.I.  Akhiezer, A.A.  Isayev, S.V.  Peletminsky,
A.P. Rekalo, and A.A. Yatsenko,   JETP {\bf 85}, 1 (1997).
\bibitem{LL5} L.D. Landau and E.M. Lifshiz,
Statistical Physics, Part 1, Pergamon press, 1980.
\bibitem{BE}
G. F. Bertsch and H. Esbensen, Ann. Phys. (NY) {\bf 209}, 327
(1991).
\bibitem{JM} J. Margueron, private communication.
\bibitem{MSH2}J. Margueron, H. Sagawa, and K. Hagino,  Preprint
arXiv:~0712.3644v1.
\bibitem{I} A.A. Isayev,  JETP Letters {\bf 82}, 551 (2005).
\end{thebibliography}
\end{document}